\documentclass[runningheads]{llncs}

\usepackage{enumitem}
\usepackage{amsmath}
\usepackage{algpseudocode}
\usepackage{algorithm}

\usepackage[T1]{fontenc}

\usepackage[labelfont=bf,font=small]{caption}
\usepackage[labelfont=bf,font=small]{subcaption}
\usepackage{graphicx}
\usepackage{hyperref}

\usepackage[separate-uncertainty=true]{siunitx}
\usepackage{booktabs,multirow}
\begin{document}
\title{MiniAnDE: a reduced AnDE ensemble to deal with microarray data}
%
%
\author{Pablo Torrijos\inst{1,2}\orcidID{0000-0002-8395-3848} \and
Jos\'e A. G\'amez\inst{1,2}\orcidID{0000-0003-1188-1117} \and
Jos\'e M. Puerta\inst{1,2}\orcidID{0000-0002-9164-5191}}
\authorrunning{P. Torrijos, JA G\'amez and JM Puerta}
%
\institute{Instituto de Investigaci\'on en Inform\'atica de Albacete (I3A). Universidad de Castilla-La Mancha. Albacete, 02071, Spain. \and
Departamento de Sistemas Inform\'aticos. Universidad de Castilla-La Mancha. Albacete, 02071, Spain.\\
\email{\{Pablo.Torrijos,Jose.Gamez,Jose.Puerta\}@uclm.es}}
\maketitle              
\begin{abstract}
This article focuses on the supervised classification of datasets with a large number of variables and a small number of instances. This is the case, for example, for microarray data sets commonly used in bioinformatics. Complex classifiers that require estimating statistics over many variables are not suitable for this type of data. Probabilistic classifiers with low-order probability tables, e.g. NB and AODE, are good alternatives for dealing with this type of data. AODE usually improves NB in accuracy, but suffers from high spatial complexity since $k$ models, each with $n+1$ variables, are included in the AODE ensemble. In this paper, we propose MiniAnDE, an algorithm that includes only a small number of heterogeneous base classifiers in the ensemble, i.e., each model only includes a different subset of the $k$ predictive variables. Experimental evaluation shows that using MiniAnDE classifiers on microarray data is feasible and outperforms NB and other ensembles such as bagging and random forest.

\end{abstract}
%
%
%
%
%
\section{Introduction}

Supervised classification, i.e. predicting the category $c \in dom(C) = \{c_1, \dots, c_r\}$ for an object $\mathbf{x}$ defined over a set of attributes $\mathbf{X} = \{X_1, \dots, X_k\}$, is one of the most profusely tackled tasks in machine learning. The objective is to learn a classifier ${\cal{C}}: X_1 \times \dots \times X_k \rightarrow C$, from a data set $\mathbf{D} = \{(\mathbf{x}^{(i)},c^{(i)})\}_{i=1}^m$, such that ${\cal{C}}$ generalises well to new data.\

In this paper we focus on a particular niche of supervised classification problems: data defined over a large number of features/attributes and with a scarce number of instances. Such data sets, where $k \gg m$, are common in microarray data problems \cite{AbdElnaby2021}, where the expression level of thousands of genes is analysed simultaneously. Still, due to the cost of obtaining samples, only a few dozen or a few hundred cases are available. This scarcity of cases means that models that need to estimate complex statistics, e.g. higher-order statistics, or measures subject to a particular context (e.g. a deep branch in a decision tree) cannot be reliably learned. A common solution to combat this curse of dimensionality is to perform a prior feature selection process \cite{BolnCanedo2014}. However, in this paper we focus on a different solution: using models that, while overall may be complex, only require estimating statistics on a very small number of variables.

\begin{figure}[htb]
\centering
    \begin{minipage}{.52\textwidth}\centering
          \includegraphics[width=\linewidth]{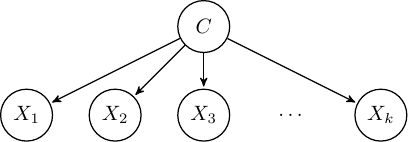}
          \captionof{figure}{Graphical structure of NB}
          \label{fig:NB}
    \end{minipage}%
    \hfill\begin{minipage}{.44\textwidth}\centering
          \includegraphics[width=0.93\linewidth]{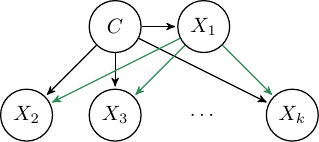}
          \captionof{figure}{Graphical structure of SPODE}
          \label{fig:SPODE}
    \end{minipage}
\end{figure}

The NB classifier \cite{webb_naive_2010} is the simplest Bayesian network model used for classification. It is based on the hypothesis (assumption) that all the predictive attributes are independent of each other given the value of the class variable (Figure \ref{fig:NB}). This independence hypothesis gives rise to the following factorisation:
\begin{equation}\label{eq:nb}
P(c,x_1,\dots,x_k) = P(c) \prod_{i=1}^{k} P(x_i | c),
\end{equation}
which enables: (1) NB does not require structural learning; (2) parametric learning is very efficient (a single pass through the BD); and (3) it is only necessary to estimate bi-variate statistics, so a small number of cases is enough.

Among the different improvements made to NB trying to circumvent the independence hypothesis, one of the most outstanding for its exceptional performance is AODE \cite{Webb2005}. AODE can be seen as an ensemble formed by $n$ SPODE (Super Parent One Dependence Estimator) classifiers, i.e. a NB extended with one attribute also being the parent of the other features (Figure \ref{fig:SPODE}). Thus, in a SPODE each variable depends on another variable apart from the class, which combined with the fact that AODE includes all the $n$ possible SPODEs, allows AODE to consider a large number of possible dependencies between attributes. Despite the strong relaxation of the NB independence assumption that AODE implies, parametric learning is still very efficient and only requires estimating three-variate statistics, so the number of cases needed remains moderate. More dependencies are considered in AnDE \cite{Webb2011}, where $n$ features play the role of super-parents in each member (SPnDE) of the ensemble. AnDE ($n\geq 2$) can manage more complex dependency relations than AODE (A1DE), however also a greater number of cases is necessary to obtain reliable estimations for $(n+1)-ary$
statistics. 

The motivation for this work comes from the fact that when dealing with microarray data, the main problem related to AnDE, even with $n=1$ (AODE), is the size of the ensemble, which can easily run out of memory. For example, let us consider a problem with $k=10000$ attributes, each taking 5 different values, as well as the class. In this case A1DE have to store 10000 SPODEs, each one with 10000 probability tables of size $5^3$, which assuming 32bits per float value means 50 GB. Of course, things are worse if we increase $n$, giving rise to the problem of dealing with \textit{big models} \cite{Arias_KBS_2017}. 

In this work we propose \textit{MiniAnDE}, an algorithm that tries to build small AnDE models in which only a subset of SPnDEs are included in the ensemble, also limiting to a subset of $\mathbf{X}$ the features included in each SPnDE. To do this, we introduce a structural learning stage in which relevant feature-class and feature-feature relations are identified. In the second stage, SPnDEs are constructed on the basis of the identified relevant relations. Experiments over nineteen microarray datasets confirm the competitiveness of our approach. 

This paper is organized as follows. Section \ref{sec:AnDE} revises the algorithm we took as our baseline, Averaged n-Dependence Estimators \cite{Webb2011}. Section \ref{sec:mAnDE} introduces the MiniAnDE classifier proposed in this paper. Section \ref{sec:experiments} presents the experimental evaluation carried out. Finally, Section \ref{sec:conclusions} concludes the paper and outlines potential avenues for future research.

%
%
\section{Averaged $n$-Dependence Estimators (AnDE)} \label{sec:AnDE}

Averaged n-Dependence Estimators (AnDE) \cite{Webb2011} extend the AODE (A1DE) algorithm by allowing $n$ super-parent variables in each model (SPnDE). As $n$ grows, the classifier estimates probability distributions of higher dimension, thus reducing its bias but probably increasing its variance, which however will be reduced when all the predictions of the base models are aggregated by the ensemble.

The class label $c^*$ of an instance $\mathbf{x}$ is obtained by:
\begin{equation}\label{eq:ande}
c^* = \arg \max_{c_i \in dom(C)} P(c_i,\mathbf{x}) = \sum_{S \in \binom{\mathbf{X}}{n}} P(c_i, \mathbf{x}_S) \prod_{X_j \in \mathbf{X}-S} P(x_j | c_i, \mathbf{x}_S),
\end{equation}
where $\binom{\mathbf{X}}{n}$ represents the subsets of $\mathbf{X}$ having exactly $n$ variables; $\mathbf{x}_S$ is the projection of $\mathbf{x}$ over $S$; the expression inside the summation is the factorization of the joint probability carried out by the SPnDE; and the summation stand for the aggregation carried out in the AnDE ensemble.

In particular, for A1DE, the previous expression reduces to:

\begin{equation}\label{eq:a1de}
c^* = \arg \max_{c_i \in dom(C)} P(c_i,\mathbf{x}) = \sum_{l=1}^{k} P(c_i, \mathbf{x}_l) \prod_{j \neq l} P(x_j | c_i, \mathbf{x}_l).
\end{equation}

The main problem in AnDE is due to its spatial complexity and the increase in the number of samples needed to make reliable estimates of increasingly larger statistics.  Thus, A1DE requires $n$ models, each with $k-1$ distributions of order 3; A2DE requires $O(n^2)$ models each with $k-2$ distributions of order 4; A3DE requires $O(n^3)$ models, each with $k-3$ distributions of order 5; etc. This means that in practice, AnDE can only be used with $n=1$ for moderate/large domains and with $n=2$ for small domains.

In literature, we can find different approaches to make AnDE usable when $n$ and/or $k$ grows. In \cite{Ana_HODE_2009}, the A1DE ensemble is replaced by a single model whose super-parent is a latent variable which is estimated by using the EM algorithm. SAnDE \cite{Chen_SAnDE_2017} and SASAnDE \cite{Chen_SASAnDE_2017} follow a model selection-based approach, which relies on the assumption that the conditional mutual information of the super parent set of attributes given the class is a good approximation of the resulting SPnDE performance. However, the study conducted in \cite{Arias_PGM18} over 43 datasets challenges this assumption and the usefulness of using mutual information-based model selection in the AnDE ensemble.

%
%
\section{MiniAnDE} \label{sec:mAnDE}
The main objective of the \textit{MiniAnDE} classifier is to reduce the enormous spatial complexity of AnDE which, in practice, impedes their use in databases with thousands of variables ($k$) in the case of A1DE and hundreds in the case of A2DE. The aim is to reduce both the number of SPnDEs generated ($s$) and the number of variables included in each SPnDE ($r_i$) so that $s \ll k$ and $r_i \ll k$. Thus, we create much smaller and faster models that can handle high-dimensional datasets.

As in \cite{Chen_SAnDE_2017}, we need to select the variable(s) that will act as super-parent(s) and thus give rise to the SPnDEs included in the AnDE model. In addition, we also have to select the {\em child} features to be included in each SPnDE. Unlike previous work, instead of calculating information-based measures, we propose to use a different machine learning model, a decision tree, from which the relationships between features can be borrowed for our MiniAnDE model. 

The use of decision trees (DTs) to select the relevant variables for a classification problem is quite old \cite{Cardie_1993}. From a probabilistic point of view, the subset of variables appearing in the tree could be seen to constitute the Markov blanket of the class variable, i.e. the set of variables that makes the rest irrelevant for classification purposes. Later, ensemble-based methods, in particular random forests, have also been used to obtain the importance of predictive variables in the classification process, using so-called out-of-bag estimation \cite{Breiman2001}. This technique has become very popular and can be found in almost any ML software, e.g. Scikit-Learn. 

In this paper we propose to use an ensemble of DTs to identify the SPnDEs to be included in our MiniAnDE model. In addition to the ability of the DTs to select the relevant variables for the class, we will also exploit the location in which these variables are placed in the tree. Thus, it is well known that one of the advantages of DTs is their context-based analysis of the data, where by context we mean a (partial) branch of the three. Therefore, we traverse the tree to identify all paths of length $n$ and create an SPnDE for each of them by setting the variables in the path as super-parents. Then, all variables in the tree that are adjacent to the super-parent variables are included as children in that SPnDE. 
To obtain a more robust MiniAnDE model we consider a set of diverse DTs, that is, an ensemble.

\begin{algorithm}[htbp]
\caption{MiniAnDE}\label{alg:miniAnDE}
\begin{algorithmic}[1]
\Require Dataset $\mathbf{D}$ defined over $\mathbf{X} \cup \{C\}$; $n$; $t$
\State $SP \leftarrow \emptyset$
\State ${\cal{T}} \leftarrow \emptyset$
\For{$i \leftarrow 1$ to $t$}
    \State $T\leftarrow $ learn a DT from a sample of $\mathbf{D}$
    \State ${\cal{T}} \leftarrow {\cal{T}} \cup \{ T \}$
    \State $SP^t \leftarrow \{$sets of $n$ consecutive variables in $T \}$
    \State $SP \leftarrow SP \cup SP^t$
\EndFor


\State $\forall sp \in SP$, children$(sp) \leftarrow \bigcup_{T \in {\cal{T}} \wedge sp \in T} \left \{ \bigcup_{X \in sp} \textrm{adjacent}(X,T)  \right \}$

\State ${\cal{M}} \leftarrow \emptyset$

\For{each $sp \in SP$ do}
    \State Create an SPnDE $m$ with $sp$ as super-parent and children$(sp)$ as features
    \State ${\cal{M}} \leftarrow {\cal{M}} \cup \{m\}$
\EndFor
\State {\bf return} ${\cal{M}}$
\end{algorithmic}
\end{algorithm}

\begin{figure}[htb]
\centering
    \begin{minipage}{.4\textwidth}\centering
          \includegraphics[width=\linewidth]{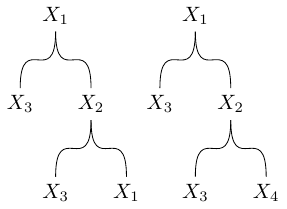}
          \subcaption{Ensemble with 2 decision trees: $T_1$ (left) and $T_2$ (right)}
          \label{subfig:arbol}
    \end{minipage}%
    \hfill\begin{minipage}{.60\textwidth}\centering
          \includegraphics[width=0.70\linewidth]{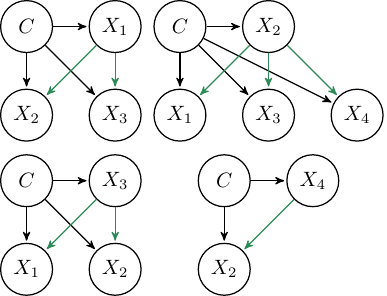}
          \subcaption{Generated SP1DEs}
          \label{subfig:SPODEs-generados}
    \end{minipage}
    \caption{MiniA1DE obtained from the ensemble $\{T_1, T_2\}$} \label{fig:ejemplomAODE}
\end{figure}
\begin{figure}[htb]
    \centering
    \includegraphics[width=\linewidth]{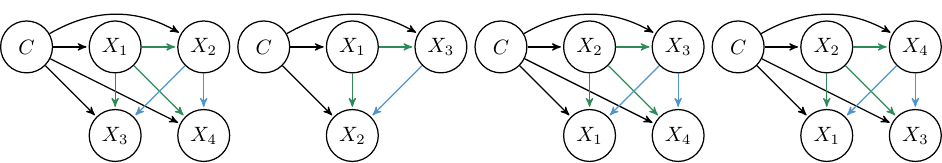}
    \caption{MiniA2DE obtained from the ensemble $\{T_1, T_2\}$ in Fig. \ref{subfig:arbol}} \label{fig:ejemplomA2DE}
\end{figure}

Algorithm \ref{alg:miniAnDE} provides a scheme of the previous idea. Let us illustrate its working process with an example taking $n=1$ and $t=2$. Let us also assume that Figure \ref{subfig:arbol} shows two DTs learnt from two different samples of $\mathbf{D}$. The algorithm starts with $T_1$ and identify $SP^1 = \{ \{X_1\}, \{X_2\}, \{X_3\}\}$. Now $SP \leftarrow SP^1$ and $T_2$ is considered. The algorithm computes $SP^2 = \{ \{X_1\}, \{X_2\}, \{X_3\}, \{X_4\} \}$, and so $SP = \{ \{X_1\}, \{X_2\}, \{X_3\}, \{X_4\}\}$. Next, children sets are computed as: children$(\{X_1\}) = \{X_2,X_3\}$, children$(\{X_2\}) = \{X_1,X_3,X_4\}$, children$(\{X_3\}) = \{X_1,X_2\}$ and children($\{X_4\}$) = $\{X_2\}$. Therefore, the SP1DEs included in the resulting MiniA1DE are those shown in Figure \ref{subfig:SPODEs-generados}. If the same process is applied with $n=2$, $SP^1 = \{ \{X_1,X_2\}, \{X_1,X_3\}, \{ \{X_2,X_3\} \}$, $SP^2 = \{ \{X_1,X_2\},$ $ \{X_1,X_3\}, \{ \{X_2,X_3\}, \{ \{X_2,X_4\} \}$ and $SP = \{ \{X_1,X_2\}, \; \{X_1,X_3\}, \; \{X_2,X_3\},$ 
$\{X_2,X_4\} \}$. Next, children sets are computed as: children$(\{X_1,X_2\}) = \{X_3,X_4\}$, children$(\{X_1,X_3\}) = \{X_2\}$, children$(\{X_2,X_3\}) = \{X_1,X_4\}$ and children$(\{X_2,X_4\}) = \{X_1, X_3\}$. Figure \ref{fig:ejemplomA2DE} shows the resulting MiniA2DE.


Like the original AnDE algorithm, MiniAnDE only works with discrete variables, so if numerical predictive attributes are included in the dataset, they must first be discretized. Once the SPnDEs have been determined, only parametric learning is required, which can be performed in a single pass through the dataset. Therefore, the complexity of learning a MiniAnDE model is dominated by the learning process of the set of decision trees. In this sense, it is worth noting that due to the small number of instances in the microarray data, the obtained tree will be shallow, which coupled with the use of only discrete (discretized) variables, results in a fast learning process. On the other hand, inference is also faster than in the original AnDE models, since only a few SPnDEs are aggregated instead of $k$.

The MiniAnDE algorithm can be instantiated with any decision tree and ensemble learning algorithm, e.g. bagging \cite{Breiman1996} or random forest \cite{Breiman2001}. This fact together with the own DT/ensemble learning hyperparameters (pruning or no-pruning, max depth, number of trees, etc.) provides a wide range of combinations to generate the MiniAnDE classifier, making possible to fine-tuning it for a given dataset.


To conclude this section, we present a possible extension of the MiniAnDE algorithm. As with AnDE, MiniAnDE is expected to be a better estimator than NB for posterior class labels probabilities. However, in some cases it is possible that some attribute configurations and class values may be missing or underrepresented in the learning dataset, resulting in a nearly uniform posterior probability distribution for the class given the input instance. To alleviate this drawback, we produce the output as a convex combination of MiniAnDE and NB, adding it to the ensemble according to a parameter $\alpha \in [0,1]$: $p(c|\mathbf{x}) = \alpha \, p_{NB}(c|\mathbf{x}) + (1-\alpha) \, p_{MiniAnDE}(c|\mathbf{x})$. We compare the MiniAnDE algorithm with $\alpha=0$ and $\alpha \neq 0$ in the experiments performed in Section \ref{sec:experiments}.

%
%
\section{Experimental evaluation} \label{sec:experiments}
In the next sections we describe the datasets utilized, the algorithms evaluated, the methodology employed, and analyze the results obtained.

\subsection{Data sets} \label{subsec:datasets}
Table \ref{tab:datasets} describes the 19 microarray data sets used to evaluate the proposed algorithms, commonly used in the literature \cite{AbdElnaby2021,BolnCanedo2014,DazUriarte2006,Zhu2007}. 

\begin{table}[htbp]
\caption{Data sets used in the experimental evaluation. \textsc{I} is the number of instances, \textsc{N} the number of predictible variables and \textsc{K} the number of classes.}\label{tab:datasets}
\resizebox{\textwidth}{!} {%
\begin{tabular*}{0.5\textwidth}{@{\extracolsep{\fill}}lS[table-format=3.0]S[table-format=6.0]S[table-format=2.0]}
\toprule
\multicolumn{1}{c}{\multirow{2}{*}{\textsc{\bfseries Data Set}}} &\multicolumn{3}{c}{\textsc{\bfseries Features}} \\
\cmidrule(){2-4}
& \multicolumn{1}{r}{\textsc{i}} & \multicolumn{1}{r}{\textsc{n}} & \multicolumn{1}{r}{\textsc{k}}\\
\midrule
\textsc{9 Tumors}        & 60	& 5726 & 9 \\
\textsc{11 Tumors}       & 174	& 12533 & 11 \\
\textsc{Breast}          & 97	& 24481 & 2 \\
\textsc{CNS}             & 60	& 7130 & 2 \\
\textsc{Colon}           & 62	& 2000 & 2 \\
\textsc{DLBCL}           & 77	& 5469 & 2 \\
\textsc{GLI}             & 85	& 22283 & 2 \\
\textsc{Leukemia}        & 72	& 7129 & 2 \\
\textsc{Leukemia 3}      & 72	& 7129 & 3 \\
\textsc{Leukemia 4}      & 72	& 7129 & 4 \\
\bottomrule
\end{tabular*}

\begin{tabular*}{0.5\textwidth}{@{\extracolsep{\fill}}lS[table-format=3.0]S[table-format=6.0]S[table-format=2.0]}
\toprule
\multicolumn{1}{c}{\multirow{2}{*}{\textsc{\bfseries Data Set}}} &\multicolumn{3}{c}{\textsc{\bfseries Features}} \\
\cmidrule(){2-4}
& \multicolumn{1}{r}{\textsc{i}} & \multicolumn{1}{r}{\textsc{n}} & \multicolumn{1}{r}{\textsc{k}}\\
\midrule
\textsc{Lung}            & 203	& 12600 & 5 \\
\textsc{Lymphoma 3}      & 66	& 4026 & 3 \\
\textsc{Lymphoma 9}      & 96	& 4026 & 9 \\
\textsc{Lymphoma 11}     & 96	& 4026 & 11 \\
\textsc{MLL}             & 72	& 12582 & 3 \\
\textsc{Ovarian}         & 253	& 15154 & 2 \\
\textsc{Prostate}        & 102	& 12600 & 2 \\
\textsc{SMK}             & 187	& 19993 & 2 \\
\textsc{SRBCT}           & 83	& 2308 & 4 \\
           & 	&  &  \\
\bottomrule
\end{tabular*}
}
\end{table}


\subsection{Reproducibility} \label{subsec:reproducibility}
The entire MiniAnDE algorithm's family has been programmed from scratch using Java (OpenJDK 8) and the library WEKA 3.9.6 \footnote{\href{https://www.cs.waikato.ac.nz/ml/weka/}{https://www.cs.waikato.ac.nz/ml/weka/'}}. All experiments were conducted on machines running the CentOS 7 operating system with an Intel Xeon E5-2650 8-Core Processor limited to 8 threads and 32 GB of RAM per execution.

To reproduce the experiments, all of the code and the execution scripts are provided at GitHub \footnote{\href{https://github.com/ptorrijos99/mAnDE}{https://github.com/ptorrijos99/mAnDE}}. 
Regarding the data, for convenience, we provide in OpenML\footnote{\href{https://www.openml.org/search?type=data&uploader\_id=\%3D\_33148}{https://www.openml.org/search?type=data\&uploader\_id=\%3D\_33148}} a common source repository for the 19 datasets, with reference to the original articles.

\subsection{Algorithms} \label{subsec:algorithms}
In this study, the following algorithms have been evaluated: 
\begin{itemize}[itemsep=5pt]
    \item The MiniAnDE algorithm introduced in Section \ref{sec:mAnDE}, with $n=1$ and $n=2$. The following parameters have been fine-tuned by using grid-search for each dataset:
    
    \begin{itemize}[itemsep=5pt]
        \item Bagging is considered to generate the ensemble of trees used to learn the structure of those SPnDEs included in the MiniAnDE model. The number of trees is taken from the set $\{50, 100, 150, 200\}$.
        \item The weight of NB is chosen from the set $\alpha = 0.02, 0.05, 0.1, 0.15, 0.2, 0.25, 0.3, 0.35, 0.4\}$. The case of $\alpha=0$ is always reported, as it corresponds to the canonical MiniAnDE as introduced in Algorithm \ref{alg:miniAnDE}. 
    \end{itemize}
    \item The Naive Bayes algorithm \cite{webb_naive_2010}.
    \item The Bagging ensemble algorithm \cite{Breiman1996}. The number of trees (50, 100, 150 and 200) is selected for each dataset by using grid-search.
    \item The Random Forest algorithm \cite{Breiman2001}. Default value $\sqrt{k}$ is used to select the random subset of variables evaluated at each split. The number of trees (50, 100, 150 and 200) is selected for each dataset by using grid-search.
\end{itemize}

Please, note that original AnDE algorithm \cite{Webb2011} is not included because of its spatial complexity. In fact, under the resources described in previous section, A1DE algorithm only can cope with 1 out of the 19 datasets (\texttt{colon}), obtaining an accuracy of 80.64. 

\subsection{Methodology} \label{subsec:methodology}
We have taken the following design decisions:

\begin{itemize}[itemsep=5pt]
    \item Each algorithm has been evaluated employing a double cross-validation. Leave-one-out cross-validation has been used for external validation, and stratified 5-fold cross-validation has been used for the internal validation in which the best hyperparameter(s) value(s) are selected by using grid-search. This approach ensured that the results were robust and not influenced by the specific partitioning of the data, especially given the small number of instances in microarray data.
    

    \item Numerical variables are discretized. Discretization intervals are learn from the training partition and then applied over the validation/test one. We used the following procedure: (1) supervised entropy-based discretization following Fayyad and Irani algorithm \cite{Fayyad1993} was applied; and (2) those variables left in a single interval are then discretized into 2 intervals (bins) by using unsupervised equal frequency. Note that variables discretized in a single bin by Fayyad and Irani algorithm are those \textit{marginally} independent to the class, but can be relevant to the class when used in conjunction with other attributes (e.g. as in an X-OR dataset).

    \item The study's results have been analyzed using the methodology specified in \cite{exreport-fuente1,exreport-fuente2}, and the analysis has been conducted using the \texttt{exreport} R package \cite{exreport}. The analysis begins by performing a Friedman test \cite{Friedman1940} with the null hypothesis that all algorithms have equal performance. If the null hypothesis is rejected, a posthoc test using Holm's procedure \cite{Holm1979} is carried out to compare all algorithms against the one ranked first by the Friedman test. Both assessments are conducted at a significance level of 5\%.
\end{itemize}

\subsection{Results} \label{subsec:results}


The summary of the accuracy results is shown in Table \ref{tab:resultados-score}, including the result of each algorithm\footnote{{\sc mAnDE} denotes the canonical MiniAnDE algorithm ($\alpha=0$) and {\sc mAnDE $\alpha>0$} denotes its combination with NB using $\alpha>0$, the parameter $\alpha$ is set using a grid search and CV, as noted above.} for each database as well as the total average of each algorithm. The algorithm(s) with the highest accuracy are highlighted in bold. In accordance with the procedure described in Section \ref{subsec:methodology}, we analyzed the results of our experiments. We found evidence to reject the null hypothesis of equal performance across all algorithms with a computed p-value of \num{1.490e-02}. The detailed results of the posthoc test are presented in Table \ref{tab:rankScore}, which shows the ranking generated by the Friedman test and the p-value adjusted using Holm's procedure (non-rejected null hypotheses are boldfaced), along with the number of wins, ties, and losses for each algorithm versus the algorithm that ranked first. Based on the statistical analysis, we draw the following conclusions:
 
\begin{table}[ht]
\caption{Accuracy of each algorithm.}\label{tab:resultados-score}
\resizebox{\linewidth}{!} {%
\begin{tabular*}{1.1\textwidth}{@{\extracolsep{\fill}}lS[table-format=2.2]S[table-format=2.2]S[table-format=2.2]S[table-format=2.2]S[table-format=3.2]S[table-format=2.2]S[table-format=3.2]}
\toprule
\multicolumn{1}{c}{\multirow{2}{*}{\textsc{\bfseries Data Set}}} &\multicolumn{7}{c}{\textsc{\bfseries Algorithm}} \\
\cmidrule(){2-8}
& \multicolumn{1}{r}{\textsc{mA1DE}} & \multicolumn{1}{r}{\textsc{mA2DE}} & \multicolumn{1}{r}{\textsc{mA1DE $\alpha>0$}} & \multicolumn{1}{r}{\textsc{mA2DE $\alpha>0$}} & \multicolumn{1}{r}{\textsc{NB}} & \multicolumn{1}{r}{\textsc{Bagging}} & \multicolumn{1}{r}{\textsc{RF}}\\
\midrule
\textsc{11} \textsc{Tumors}   &     83.91 &     85.06 & \textbf{89.66} &        88.51 &  84.48 &    87.36 &  85.06 \\
\textsc{9} \textsc{Tumors}    &     33.33 &     35.00 &        50.00 &        48.33 & \textbf{53.33} &    36.67 &  36.67 \\
\textsc{Breast}               &     67.01 &     67.01 &        64.95 &        68.04 & \textbf{69.07} &    67.01 &  62.89 \\
\textsc{CNS}                  &     60.00 &     65.00 &        63.33 &        71.67 &  60.00 & \textbf{73.33} &  65.00 \\
\textsc{Colon}                &     85.48 & \textbf{87.10} & \textbf{87.10} & \textbf{87.10} & \textbf{87.10} &    85.48 & \textbf{87.10} \\
\textsc{DLBCL}                &     \textbf{89.61} &     84.42 &        84.42 &        81.82 &  80.52 &    87.01 &  88.31 \\
\textsc{GLI}                  &     \textbf{87.06} &     85.88 &        85.88 &        84.71 &  82.35 &    85.88 &  85.88 \\
\textsc{Leukemia}             &     95.83 &     95.83 & \textbf{97.22} &        94.44 &  87.50 &    91.67 &  94.44 \\
\textsc{Leukemia} \textsc{3}  &     94.44 &     94.44 & \textbf{95.83} &        94.44 &  83.33 &    94.44 &  87.50 \\
\textsc{Leukemia} \textsc{4}  &     \textbf{91.67} &     90.28 &        90.28 &        90.28 &  79.17 &    88.89 &  77.78 \\
\textsc{Lung}                 &     90.64 &     91.13 &        92.61 &        94.09 &  72.91 & \textbf{96.55} &  89.16 \\
\textsc{Lymphoma} \textsc{11} &     77.08 &     81.25 &        90.62 & \textbf{91.67} & \textbf{91.67} &    81.25 &  84.38 \\
\textsc{Lymphoma} \textsc{3}  &     95.45 &     93.94 &        98.48 &        98.48 & \textbf{100.00} &    93.94 &  93.94 \\
\textsc{Lymphoma} \textsc{9}  &     78.12 &     76.04 &        89.58 &        91.67 & \textbf{95.83} &    81.25 &  81.25 \\
\textsc{MLL}                  &     94.44 &     95.83 &        \textbf{97.22} &        95.83 &  90.28 &    93.06 &  94.44 \\
\textsc{Ovarian}              &     97.63 & \textbf{98.42} &        97.63 &        98.02 &  92.49 &   98.02   &  95.26 \\
\textsc{Prostate}             &     \textbf{93.14} &     91.18 &        91.18 &        88.24 &  65.69 &    91.18 &  86.27 \\
\textsc{SMK}                  &     70.05 &     70.05 & \textbf{71.66} & \textbf{71.66} &  65.24 &  70.59  &  65.24 \\
\textsc{SRBCT}                &     98.80 &     97.59 &        98.80 &        97.59 &  92.77 &    95.18 & \textbf{100.00} \\
\midrule
\textsc{\bfseries Mean}       &     83.35 &     83.44 &        86.13 &        \textbf{86.14} &  80.72 &   84.15  & 82.14 \\
\bottomrule
\end{tabular*}
}
\end{table}

\begin{table}[ht]
\caption{Post-hoc test results for the accuracy of each algorithm.}\label{tab:rankScore}
\resizebox{\textwidth}{!} {%
\begin{tabular*}{1\textwidth}{@{\extracolsep{\fill}}lS[detect-weight,table-format=1.3]S[table-format=1.2]S[table-format=2.0]S[table-format=1.0]S[table-format=1.0]}
\toprule
\textsc{\bfseries Algorithm} & \textsc{\bfseries \textit{p}-value} & \textsc{\bfseries Rank} & \textsc{\bfseries Win} & \textsc{\bfseries Tie} & \textsc{\bfseries Loss}\\
\midrule
\textsc{MiniA1DE ($\alpha >0$)} & {-} & 2.84 & {-} & {-} & {-}\\
\textsc{MiniA2DE ($\alpha >0$)} & \bf \num{7.073e-01} & 3.11 & 9 & 4 & 6\\
\textsc{MiniA2DE ($\alpha=0$)} & \bf \num{2.419e-01} & 4.00 & 11 & 5 & 3\\
\textsc{Bagging} & \bf \num{2.419e-01} & 4.08 & 12 & 2 & 5\\
\textsc{MiniA1DE ($\alpha=0$)} & \bf \num{2.419e-01} & 4.16 & 12 & 2 & 5\\
\textsc{Random Forest} & \num{3.432e-02} & 4.74 & 14 & 2 & 3\\
\textsc{Naive Bayes} & \num{8.492e-03} & 5.08 & 13 & 1 & 5\\
\bottomrule
\end{tabular*}
}
\end{table}

\begin{itemize}
    \item The MiniA1DE algorithm with $\alpha>0$ is ranked in the first place, although there is no significant difference (confidence level 0.05) with respect to the other three MiniAnDE algorithms and bagging. A significant difference is observed with respect to NB and random forest. 

    \item Both MiniAnDE algorithms with $\alpha>0$ rank ahead, although without significant difference among them, of their counterpart canonical versions without incorporating NB. This corroborated the fact that in some cases, due to the small sample size in microarray datasets, it is good to incorporate the prediction of a simple low-bias classifier.



    \item Regarding the use of $n=1$ or $n=2$, there do not seem to be major differences in either MiniAnDE or MiniAnDE-NB, with either option working better depending on the data set, resulting in an almost identical average accuracy.

    \item NB is ranked in the last position, which is not unexpected due to the fact that it is by far the simpler model tried. However, it is interesting to observe the bad results obtained by RF, which is ranked behind bagging. It seems that the use of pseudorandom DTs does not match with the large number of variables and small data size of microarray data.
\end{itemize}

\begin{table}[ht]
\caption{Execution time per L.O.O. iteration (seconds) of each algorithm.}\label{tab:resultados-tiempo}
\resizebox{1\linewidth}{!} {%
\begin{tabular*}{1.1\textwidth}{@{\extracolsep{\fill}}lS[table-format=1.2]S[table-format=1.2]S[table-format=1.2]S[table-format=1.2]S[table-format=1.2]S[table-format=1.2]S[table-format=1.2]}
\toprule
\multicolumn{1}{c}{\multirow{2}{*}{\textsc{\bfseries Data Set}}} &\multicolumn{7}{c}{\textsc{\bfseries Algorithm}} \\
\cmidrule(){2-8}
& \multicolumn{1}{r}{\textsc{mA1DE}} & \multicolumn{1}{r}{\textsc{mA2DE}} & \multicolumn{1}{r}{\textsc{mA1DE $\alpha>0$}} & \multicolumn{1}{r}{\textsc{mA2DE $\alpha>0$}} & \multicolumn{1}{r}{\textsc{NB}} & \multicolumn{1}{r}{\textsc{Bagging}} & \multicolumn{1}{r}{\textsc{RF}}\\
\midrule
\textsc{11} \textsc{Tumors}   &      4.05 &      4.94 &         5.31 &         5.23 & \textbf{0.83} &     4.50 &          1.50 \\
\textsc{9} \textsc{Tumors}    &      0.96 &      0.92 &         0.95 &         1.05 & \textbf{0.20} &     0.95 &          0.37 \\
\textsc{Breast}               &      3.23 &      3.27 &         3.49 &         3.58 & \textbf{0.83} &     3.44 &          1.41 \\
\textsc{CNS}                  &      0.59 &      0.66 &         0.66 &         0.68 & \textbf{0.18} &     0.61 &          0.42 \\
\textsc{Colon}                &      0.72 &      0.74 &         0.74 &         0.77 & \textbf{0.18} &     0.66 &          \textbf{0.18} \\
\textsc{DLBCL}                &      0.40 &      0.43 &         0.54 &         0.42 & \textbf{0.14} &     0.37 &          0.28 \\
\textsc{GLI}                  &      2.08 &      2.07 &         1.94 &         2.12 & \textbf{0.66} &     1.58 &          1.03 \\
\textsc{Leukemia}             &      0.46 &      0.48 &         0.44 &         0.45 & \textbf{0.19} &     0.47 &          0.34 \\
\textsc{Leukemia} \textsc{3}  &      0.60 &      0.53 &         0.52 &         0.52 & \textbf{0.17} &     0.53 &          0.37 \\
\textsc{Leukemia} \textsc{4}  &      0.73 &      0.60 &         0.69 &         0.66 & \textbf{0.19} &     0.57 &          0.36 \\
\textsc{Lung}                 &      3.69 &      3.81 &         3.80 &         4.03 & \textbf{0.95} &     4.00 &          1.36 \\
\textsc{Lymphoma} \textsc{11} &      0.96 &      1.06 &         1.02 &         0.90 & \textbf{0.18} &     0.87 &          0.42 \\
\textsc{Lymphoma} \textsc{3}  &      0.37 &      0.35 &         0.32 &         0.33 & \textbf{0.12} &     0.28 &          0.24 \\
\textsc{Lymphoma} \textsc{9}  &      0.81 &      0.86 &         0.88 &         0.86 & \textbf{0.21} &     0.80 &          0.29 \\
\textsc{MLL}                  &      0.92 &      0.81 &         0.89 &         0.85 & \textbf{0.29} &     0.70 &          0.55 \\
\textsc{Ovarian}              &      3.00 &      3.03 &         3.07 &         2.98 & \textbf{1.23} &     2.94 &          1.50 \\
\textsc{Prostate}             &      1.38 &      1.27 &         1.38 &         1.46 & \textbf{0.41} &     1.37 &          0.71 \\
\textsc{SMK}                  &      7.42 &      8.98 &         7.69 &         7.98 & \textbf{1.22} &     7.46 &          2.07 \\
\textsc{SRBCT}                &      0.28 &      0.29 &         0.29 &         0.29 & \textbf{0.09} &     0.30 &          0.18 \\
\midrule
\textsc{\bfseries Mean}       &     1.72 &     1.85 &        1.82 &      1.85 &  \textbf{0.44} & 1.70 & 0.72 \\
\bottomrule
\end{tabular*}
}
\end{table}

As for computational efficiency, the CPU time is shown in Table \ref{tab:resultados-tiempo}. As expected, NB is the fastest algorithm (linear in the number of variables and instances). On the other hand, the MiniAnDE algorithms require an affordable amount of CPU time, almost identical to bagging, the classifier it uses to train the trees. Furthermore, the effect of using MiniAnDE with $\alpha>0$ is practically insignificant. In general, we can say that the MiniAnDE approach is the best choice among the tested hypotheses when dealing with microarray data.



%
%
\section{Conclusions} \label{sec:conclusions}

A new algorithm for learning AnDE-like classifiers has been proposed. The method is tailored to the special case of microarray data, where few data instances are available but the number of variables is so large (thousands) that standard AnDE classifiers do not fit in memory. The proposed algorithm incorporates a structural learning stage, which based on the use of shallow decision trees, allows the selection of a few SPnDEs in the resulting MiniAnDE ensemble. Furthermore, a small subset of variables is included in each SPnDE, leading to a very light model regarding spatial needs and providing fast inference. The experiments' results over 19 microarray datasets show the competitivity of our proposal regarding decision tree-based ensembles, both in accuracy and efficiency.

As future works, we plan to study our proposal without the need of discretizing numerical variables, by considering AnDE models based on the use of conditional Gaussian networks \cite{Ana_ICML_2009}.


\subsubsection{Acknowledgements} This work has been funded by the Government of Castilla-La Mancha and ``ERDF A way of making Europe'' under project SBPLY/21/180225/000062. It is also partially funded by MCIN/AEI/10.13039/501100011033 and ``ESF Investing your future'' through the projects PID2019--106758GB--C33 and FPU21/01074. 

This preprint has not undergone peer review or any post-submission improvements or corrections. The Version of Record of this contribution is published in Communications in Computer and Information Science, vol 1826, and is available online at \href{https://doi.org/10.1007/978-3-031-34204-2\_12}{https://doi.org/10.1007/978-3-031-34204-2\_12}.

%
%
%
\bibliographystyle{splncs04}
\bibliography{biblio}

\end{document}